\documentclass[preprint,showkeys,superscriptaddress,secnumarabic,amsfonts,showpacs,amsmath,amssymb]{revtex4}
\usepackage[dvips]{color}
\usepackage{epsfig}
\usepackage{amsmath}
\usepackage{graphicx}
\def\Box{\hbox{$\rlap{$\sqcup$}\sqcap$}}

\begin{document}
\title{Cosmic Dynamics in $F(R,\phi)$ Gravity}

\author{H. Farajollahi}
\email{hosseinf@guilan.ac.ir} \affiliation{Department of Physics,
University of Guilan, Rasht, Iran}
\author{M. Setare}
\email{rezakord@ipm.ir} \affiliation{Department of Science , Payam Noor University , Bijar, Iran}
\author{F. Milani}
\email{fmilani@guilan.ac.ir} \affiliation{Department of Physics,
University of Guilan, Rasht, Iran}
\author{F. Tayebi}
\email{ftayebi@guilan.ac.ir} \affiliation{Department of Physics,
University of Guilan, Rasht, Iran}

\date{\today}

\date{\today}

\begin{abstract}
\noindent \hspace{0.35cm} In this paper we consider FRW cosmology in
$F(R,\phi)$ gravity. It is shown that in particular cases the bouncing behavior may appears in
the model whereas the equation of state (EoS) parameter may crosses the phantom divider.
For the dynamical universe, quantitatively we also find parameters in the model which satisfies two
independent tests:the model independent Cosmological Redshift Drift (CRD) test and the type Ia supernova luminosity distances.

\end{abstract}

\pacs{04.50.Kd; 98.80.-k}

\keywords{Scalar-tensor theory; $F(R,\phi)$ gravity; bouncing universe; phantom crossing; velocity drift; luminosity distance}

\maketitle

\section{Introduction}

There are cosmological observations, such
as Super-Nova Ia (SNIa), measurements of the cosmic microwave background (CMB)
temperature fluctuations by the Wilkinson Microwave
Anisotropy Probe (WMAP), the large scale red-shift data from the Sloan Digital Sky Survey
(SDSS) and Chandra X-ray Observatory, that disclose some cross-checked information
of our universe. Based on these observations the universe is very nearly spatially flat,
and consists of approximately $75\%$ dark energy (DE), $25\%$ dust matter (cold dark matter plus baryons), and
negligible radiation, in which the DE may drive the cosmic acceleration expansion {\cite{Riess-A.J.607}}-{\cite{Allen-M.N.353}}.

One possible candidate for DE known as the cosmological constant which is given by the vacuum energy with a constant equation of state may justify the late time acceleration \cite{Kratochvil}--\cite{Appleby-PLB654}. However, its value is so small ( of order $10^{-33}eV$), in comparison with the Planck scale ($10^{19}GeV$) that suffer from fine-tuning and the coincidence problems \cite{Mohapatra}. Instead, there are some other DE models are made by some exotic matter like phantom (field with negative energy) or some other scalar fields \cite{Caldwell-PLB 545, setare1}. In general, there are many approaches to explain the origin of DE, which can be broadly classified into two classes
{\cite{Peebles-RMP75,Copeland-IJMPD15}}. Either, through introduction of a more or less exotic form of
matter such as phantom, quintessence {\cite{Fujii-PRD26}}
 or k-essence {\cite{Chiba-PRD62}}, or through modification of gravity such as brane-world models {\cite{Dvali-PLB485}}, Gauss-Bonnet dark energy {\cite{Milani-prd}}, $f(R)$ gravity \cite{Capozziello-IJM12}-{\cite{Nojiri-PLB659}}, scalar-tensor theories
{\cite{Uzan-PRD59}}, generalized scalar tensor theories \cite{generalised}, etc.

Furthermore, according to the above observational data there exists the possibility that the EoS parameter has a dynamical behavior, evolving from larger than -1 (non-phantom phase) to less than -1 (phantom one ), namely, crosses -1 (the phantom divide)
currently or in near future. In the framework of general relativity, a number of attempts to realize the phantom crossing have been made by many authors \cite{Milani-MPLA24}--\cite{Guo-PLB608}. In the present work, we study the crossing behavior in the generalized scalar -tensor theory of gravity, $F(R,\phi)$.

A bouncing universe which provides a possible alternative to the Big Bang singularity problem in standard cosmology has recently attracted a lot of interest in the field of string
theory and modified gravity {\cite{Biswas, Kanti}}. In bouncing cosmology, within the framework of the standard FRW cosmology, the null energy condition (NEC) for a period of time around the
bouncing point is violated . Moreover, after the bouncing when the universe enters into the hot Big Bang era, the EoS parameter
in the universe crosses the phantom divide line \cite{Cai, Milani-MPLA}. There are cosmological models in the framework of modified gravity which address both bouncing universe and crossing the barrier and again in here we discuss these in the framework of $F(R,\phi)$ gravity.

To understand the true nature of
the driving force of the accelerating universe, mapping of the cosmic expansion
is very crucial \cite{Linder}. Here, we examine two observational tests
in different redshift ranges to  explain the expansion history of
the universe \cite{Nordin, Huterer}. The first probe we investigate is "Cosmological
Redshift Drift" (CRD) test which maps the expansion of the universe
\cite{Lis, Cristiani} (For a review see \cite{Jain}--\cite{Loeb}) and measures the dynamics of the
universe directly via the Hubble expansion factor. The second probe is the observations of the luminosity distance –- redshift relation for SNIa that again verifies the late-time accelerated expansion of the universe.

This work is arranged as follows. In the next section we present the $F(R,\phi)$ gravity. We drive the field equations and investigate analytically the conditions for the EoS parameter in the model to manifest phantom crossing and numerically analysis both bouncing and crossing the phantom line. Section three is devoted to the observational tests by including them in our analysis through driving CRD test and luminosity distance for the model. We also visit the Chevallier-Polarski-Linder (CPL) parametrization model \cite{Chevalier}  in a comparison with our model and observational data. We finally conclude with a summary and remarks in section four.

\section{ The model}

We start with the scalar-tensor theory of gravity where the scalar curvature $R$ in the action is replaced by $F(R,\phi)$,
\begin{eqnarray}\label{ac}
S=\int{d^{4}x\sqrt{-g}F(R,\phi)},
\end{eqnarray}
with $g$ is the determinant of the metric tensor $g_{\mu\nu}$ and $\phi$ is a scalar field. It
may originally turns out that $F(R,\phi)$ model which may be considered as a generalisation of $F(R)$ gravity, (with or without matter) can be used to solve some of the problems in cosmology. However, in general, the construction of $F(R,\phi)$ is
not explicit and it is necessary to solve the second order
differential and algebraic equations. In here, we consider a polynomial form of $F(R,\phi)$ as $\sum_{n=0}^{\infty} P_{n}(\phi)R^{n}$, as an extension of the formulation given in \cite{Nojiri-PLB659}\cite{Nojiri-PRD68}, where the scalar field $\phi$ is without kinetic term.

The action Eq. (\ref{ac}) is then rewritten as
\begin{eqnarray}\label{ac1}
S=\int{d^{4}x\sqrt{-g}(\sum_{n=0}^{\infty} P_{n}(\phi)R^{n})}.
\end{eqnarray}
and its variation with respect to the metric tensor $g_{\mu\nu}$ gives us the field equations as
\begin{eqnarray}\label{T}
0&=&-\frac{1}{2}g_{\mu\nu}\sum_{n=0}^{\infty} P_{n}(\phi)R^{n}+\left(R_{\mu\nu}+\left(g_{\mu\nu}\Box-\nabla_{\mu}\nabla_{\nu}\right)\right)\sum_{n=0}^{\infty} nP_{n}(\phi)R^{n-1}.
\end{eqnarray}
Moreover, variation of the action (\ref{ac1}) with respect to the scalar field $\phi$ gives
\begin{eqnarray}\label{eq}
\sum_{n=0}^{\infty} P'_{n}(\phi)R^{n}=0,
\end{eqnarray}
where $P'_n(\phi)=\frac{dP_n(\phi)}{d\phi}$.
The field equations (\ref{T}) corresponding to standard spatially-flat
FRW universe for the time and spacial components of the metric yields,
\begin{eqnarray}
0&=&-3H^{2}\sum_{n=0}^{\infty} nP_{n}(\phi)R^{n-1}-3\dot{H}\sum_{n=0}^{\infty} nP_{n}(\phi)R^{n-1}+3H\frac{d}{dt}\sum_{n=0}^{\infty} nP_{n}(\phi)R^{n-1}\nonumber\\
&+&\frac{1}{2}\sum_{n=0}^{\infty} P_{n}(\phi)R^{n},\label{f1}\\
0&=&-(3H^2+2\dot{H})\sum_{n=0}^{\infty} nP_{n}(\phi)R^{n-1}+\left(\frac{d^2}{dt^2}+2H\frac{d}{dt}+\dot{H}\right)\sum_{n=0}^{\infty} nP_{n}(\phi)R^{n-1}\nonumber\\
&+&\frac{1}{2}\sum_{n=0}^{\infty} P_{n}(\phi)R^{n}.\label{f2}
\end{eqnarray}
In comparison with the standard Friedman equations,$\frac{\rho_{eff}}{M_p^2}=3H^{2}$ and $\frac{p_{eff}}{M_p^2}=-3H^{2}-2\dot{H}$
one can consider the model as standard model with the effect of the $F(R,\phi)$ gravity modification is contributed in the energy density and pressure of the Friedman equations. After some algebraic calculation, one can read the equivalently effective energy density and pressure from the above equations as,
\begin{eqnarray}
\rho_{eff}&=&\frac{-M_p^2}{\sum_{n=0}^{\infty} nP_{n}(\phi)R^{n-1}}\left\{3(\dot{H}-H\frac{d}{dt})\sum_{n=0}^{\infty} nP_{n}(\phi)R^{n-1}-\frac{1}{2}\sum_{n=0}^{\infty} P_{n}(\phi)R^{n}\right\},\label{rho}\\
p_{eff}&=&\frac{-M_p^2}{\sum_{n=0}^{\infty} nP_{n}(\phi)R^{n-1}}\left\{\left(\frac{d^2}{dt^2}+2H\frac{d}{dt}+\dot{H}\right)\sum_{n=0}^{\infty} nP_{n}(\phi)R^{n-1}+\frac{1}{2}\sum_{n=0}^{\infty} P_{n}(\phi)R^{n}\right\}.\label{p}
\end{eqnarray}
Using Eqs. (\ref{rho}) and (\ref{p}), the conservation equation can be read as,
\begin{eqnarray}\label{rhodot}
\dot{\rho}_{eff}+3H\rho_{eff}(1+\omega_{eff})=0,
\end{eqnarray}
where $\omega_{eff}=\frac{p_{eff}}{\rho_{eff}}$. From Eq. (\ref{f1}) one also obtain,
\begin{eqnarray}\label{h}
H=\frac{\dot{\mathcal{A}}}{2}\pm\left\{(\frac{\dot{\mathcal{A}}}{2})^2
-\dot{H}+\frac{1}{6\frac{d}{dR}\mathcal{B}}\right\}^{\frac{1}{2}},
\end{eqnarray}
where
\begin{eqnarray}
\mathcal{A}&=&\ln\sum_{n=0}^{\infty}nP_n(\phi)R^{n-1},\\
\mathcal{B}&=&\ln\sum_{n=0}^{\infty}P_n(\phi)R^{n}.
\end{eqnarray}
To study the behavior of the scale factor of the universe and Hubble parameter, for simplicity, we consider two
linear and quadratic forms of the polynomial $F(R,\phi)$. Numerically, we solve the equations (\ref{eq})-(\ref{p}) for these cases and obtain the results given in Figs. (1) and (2). As can be seen, in Fig. (1), in linear case, for exponential behavior of the $P_0(\phi)$ and power law behavior of the $P_1(\phi)$, the scale factor bounces at $t=0$. In Fig. (2), with similar exponential behavior of the $P_0(\phi)$ but in quadratic form of  $F(R,\phi)$ where $P_2(\phi)$ is power law and $P_1(\phi)$ is logarithmic, we obtain different dynamics for the scale factor. In this case the universe has an inflection at $t=0$.

\begin{tabular*}{2.5 cm}{cc}
\includegraphics[scale=.35]{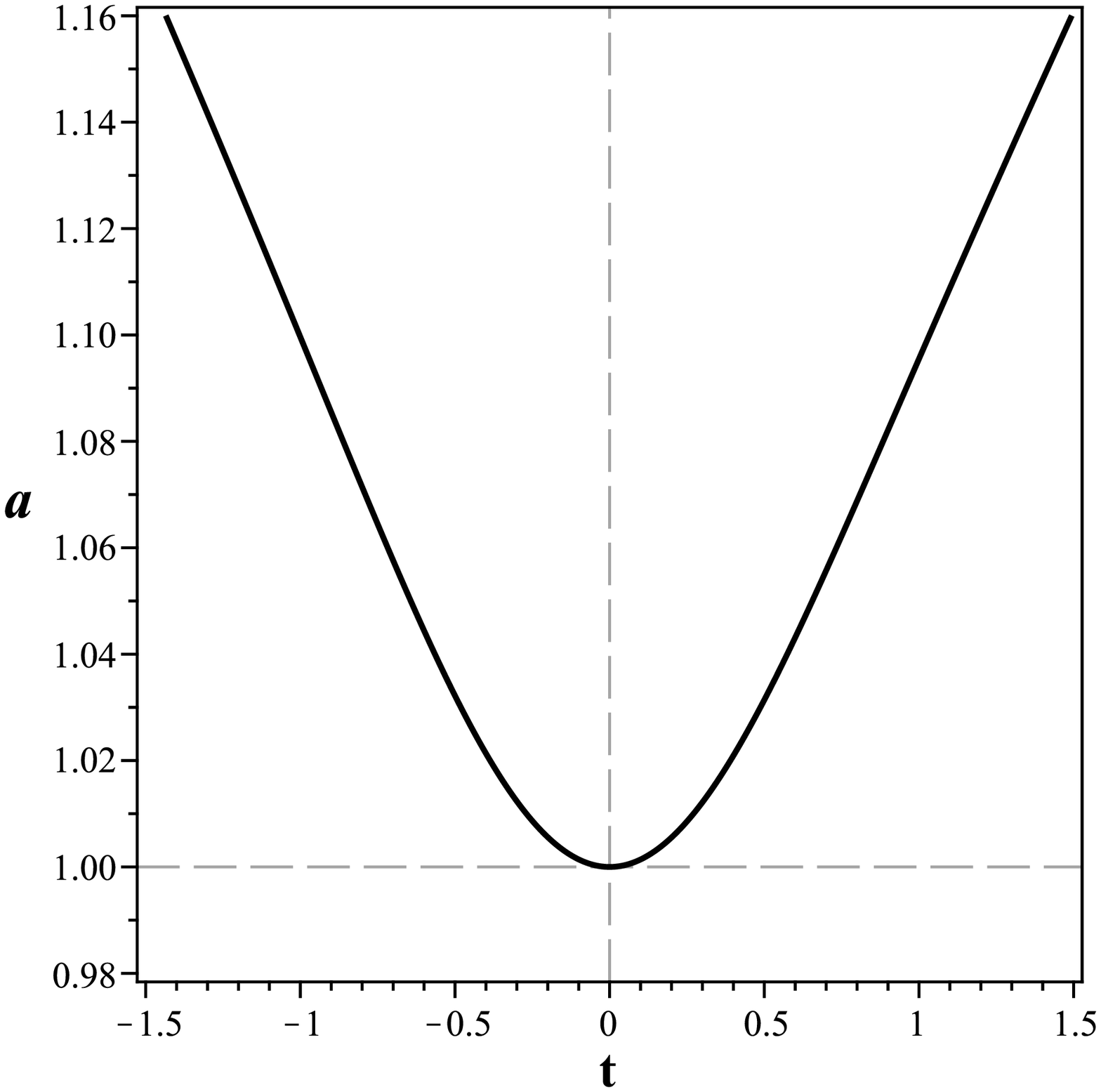}\hspace{0.2 cm}\includegraphics[scale=.35]{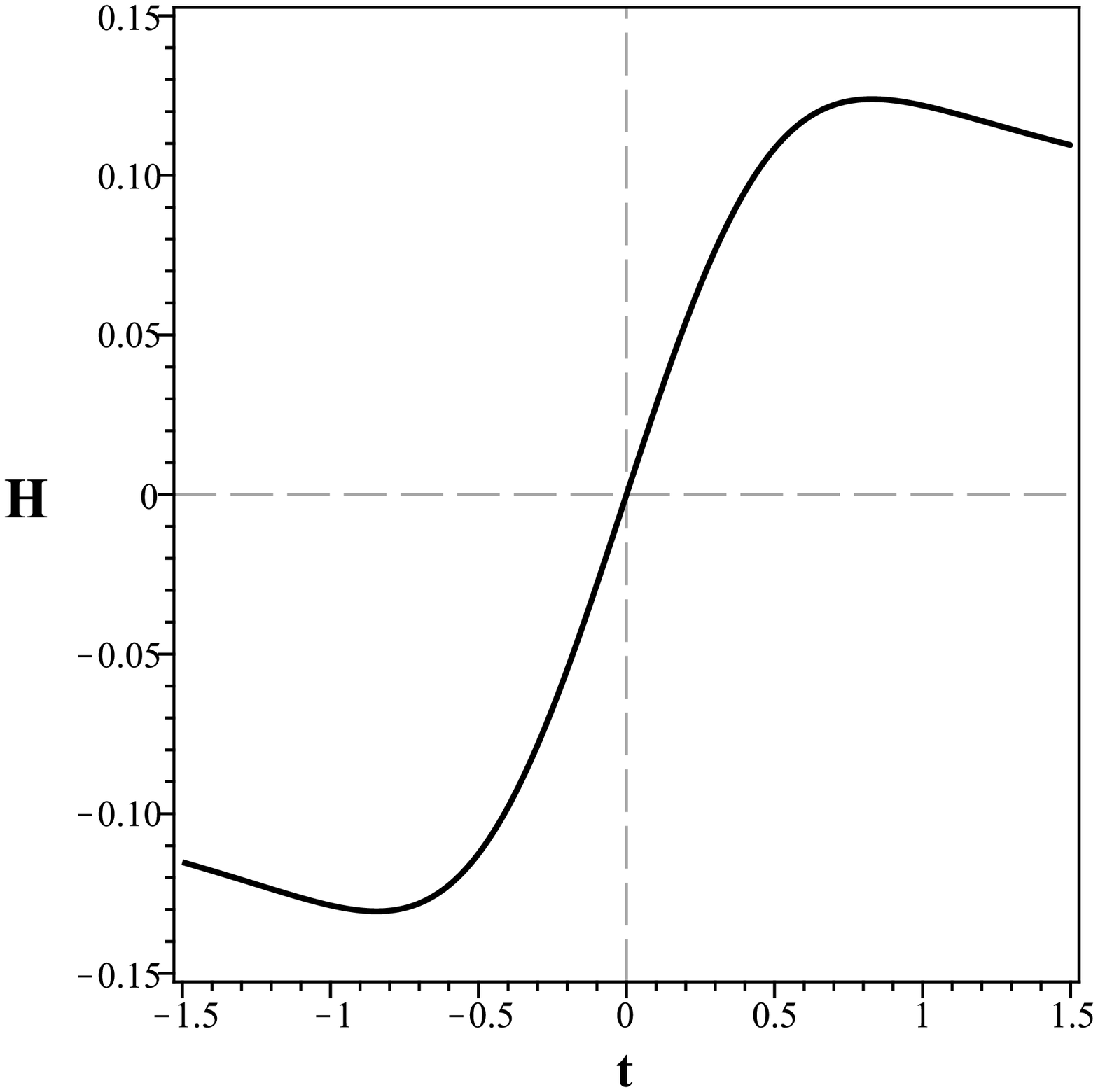}\hspace{0.2 cm}\\
\hspace{1 cm}Fig. 1: \, The graph of $a$ and $H$   plotted as functions of time for linear \\
\hspace{1 cm}  $F(R,\phi)=P_1(\phi)R+P_0(\phi)$,  $P_1(\phi)=\phi^\alpha$, $P_0(\phi)=e^{\beta \phi}$, for $\alpha = -0.125$ and $ \beta= -0.5$.\\
\hspace{1 cm}   Initial value is $\phi_0(0)=0.5$.\\
\end{tabular*}\\
\\
\begin{tabular*}{2.5 cm}{cc}
\includegraphics[scale=.35]{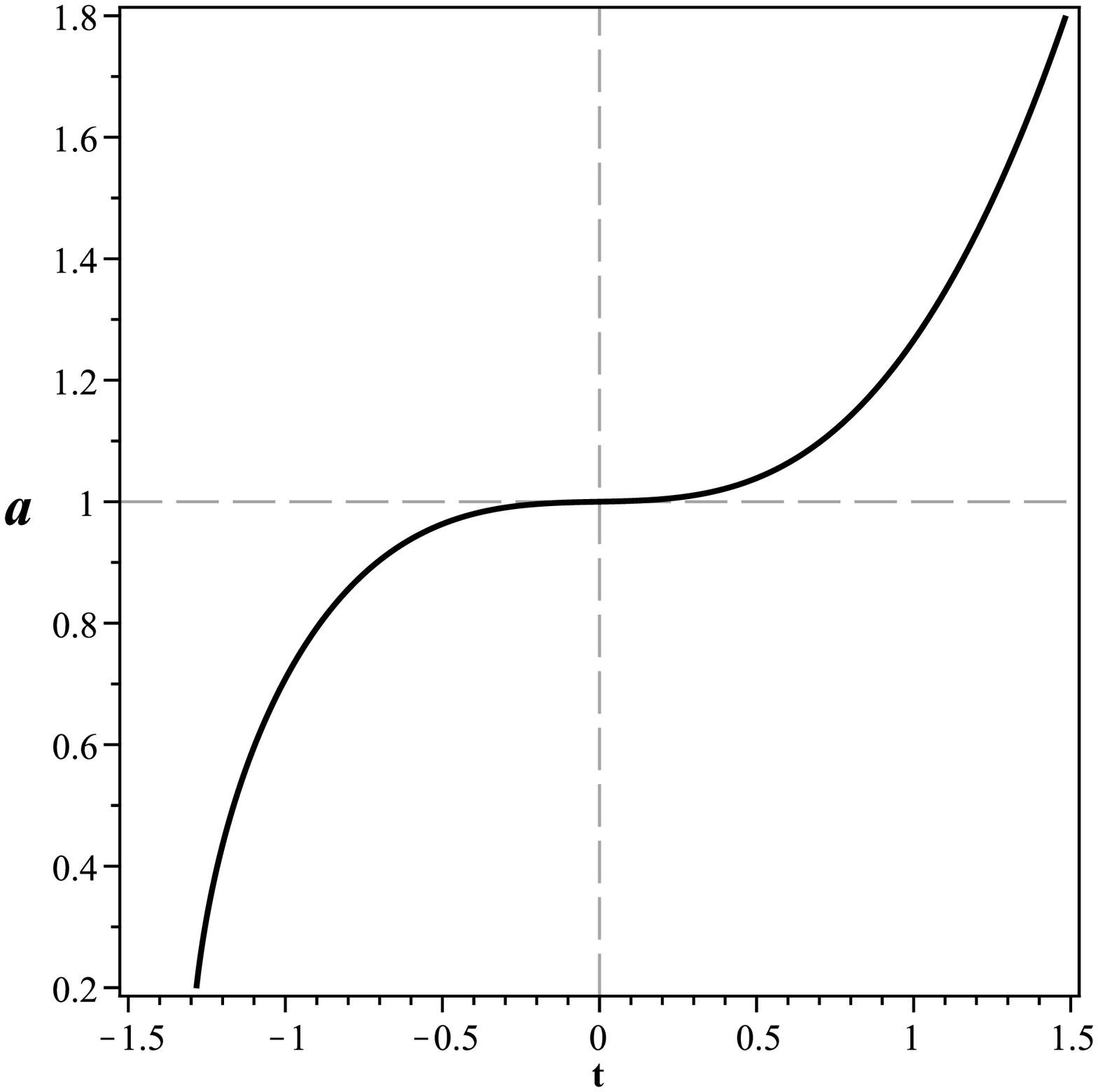}\hspace{0.2 cm}\includegraphics[scale=.35]{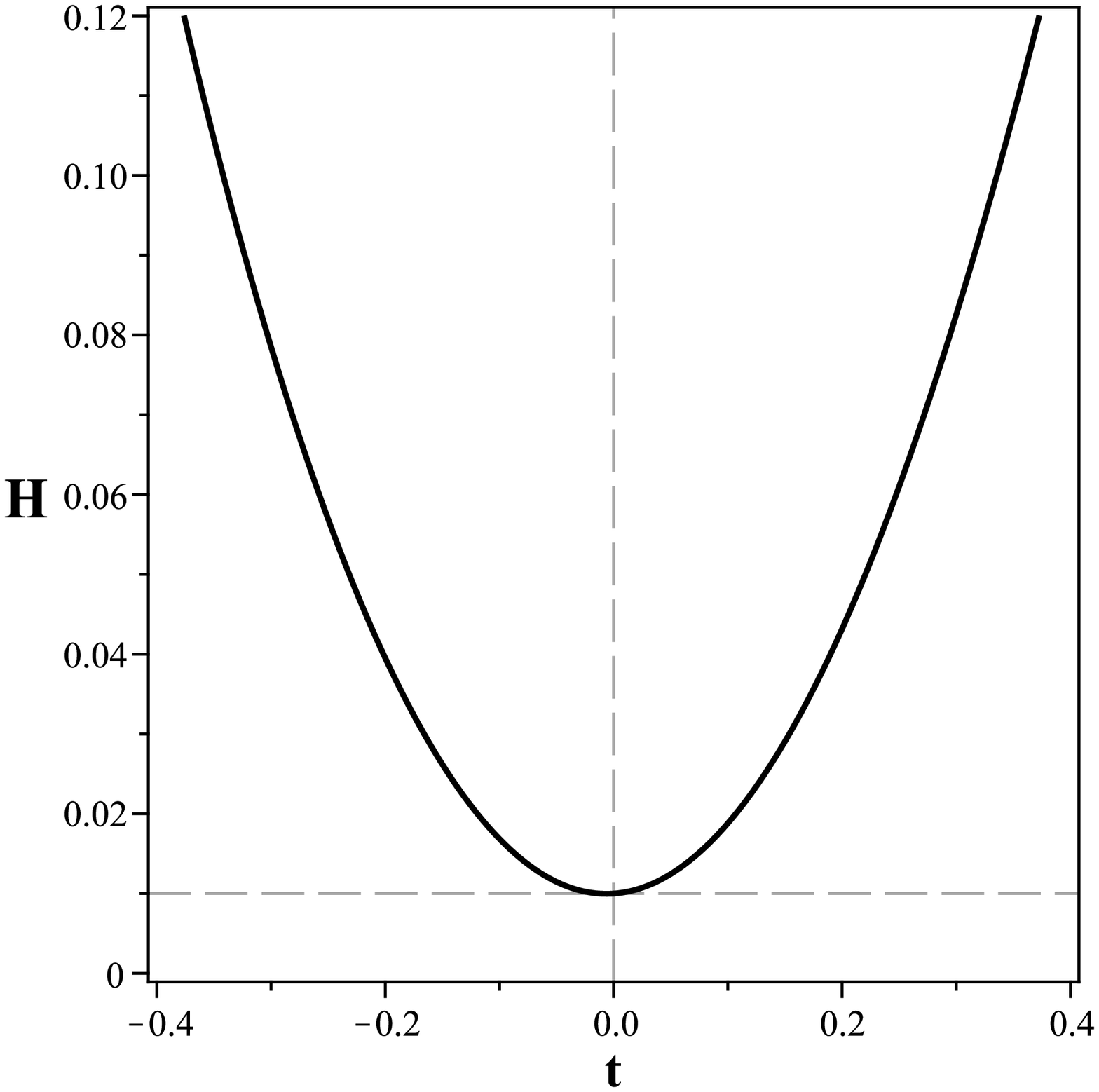}\\
\hspace{1 cm}Fig. 2: \, The graph of $a$ and $H$   plotted as functions of time for quadratic \\
\hspace{1 cm}  $F(R,\phi)=P_{2}(\phi)R^{2}+P_{1}(\phi)R+P_{0}(\phi)$, $P_{2}(\phi)=\ln(\phi)$, $P_{1}(\phi)=\phi^{\alpha}$ and $P_{0}(\phi)=e^{\beta \phi}$,\\
\hspace{1 cm}  for $\alpha = -0.125$ and $ \beta= -0.5$. Initial value is $\phi(0)=0.5$.
\end{tabular*}\\

At this stage we investigate the cosmological evolution of EoS parameter, and show that there are conditions that
cause the EoS parameter crosses the phantom divide line. Analytically, one requires $\dot{H}=-\frac{1}{2}(\rho_{eff} + p_{eff})$ vanishes for a particular $\phi_0$ and changes the sign after the crossing. One may explore this possibility by checking the condition $\frac{d}{dt}(\rho_{eff}+p_{eff})\neq 0$ when $\omega_{eff} \rightarrow -1$. From Eqs. (\ref{f1}) and (\ref{f2}), one gets,
\begin{eqnarray}\label{Hdot}
\dot{H}=-\frac{1}{2}\left\{\ddot{\mathcal{A}}+\dot{\mathcal{A}}^2-H\dot{\mathcal{A}}\right\}\cdot
\end{eqnarray}
Also we have $\frac{d}{dt}(\rho_{eff}+p_{eff})=
-2\ddot{H}$ or,
\begin{eqnarray}\label{Hddot}
\ddot{H}=-\frac{1}{2}\left\{\dddot{\mathcal{A}}+(2\dot{\mathcal{A}}-H)\ddot{\mathcal{A}}-\dot{H}\dot{\mathcal{A}}\right\}\cdot
\end{eqnarray}
One can find that in order to have phantom crossing one of the following conditions might be satisfied when $\omega_{eff}\rightarrow -1$:
\begin{itemize}
  \item (a) If $\ddot{\mathcal{A}}=0$ or $2\dot{\mathcal{A}}=H$ then $\dddot{\mathcal{A}}\neq 0$,
  \item (b) If $\dddot{\mathcal{A}}=0$ then $\ddot{\mathcal{A}}\neq 0$ and $2\dot{\mathcal{A}}\neq H$.
\end{itemize}
Numerically, As shown in Fig. (3), for both linear and quadratic $F(R,\phi)$, the EoS parameter crosses the boundary at $t>0$ in the past.

\begin{tabular*}{2.5 cm}{cc}
\includegraphics[scale=.35]{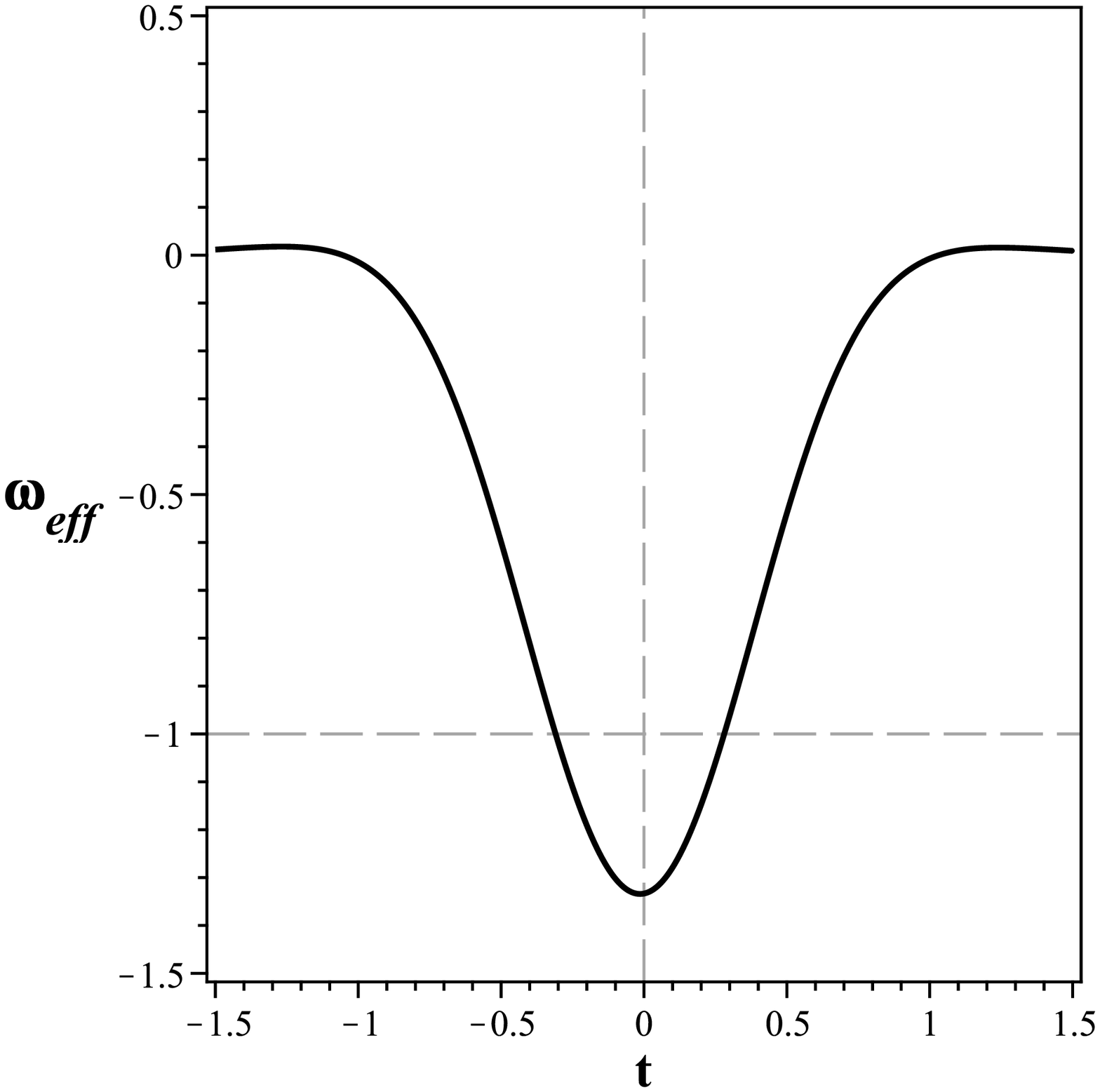}\hspace{0.2 cm}\includegraphics[scale=.35]{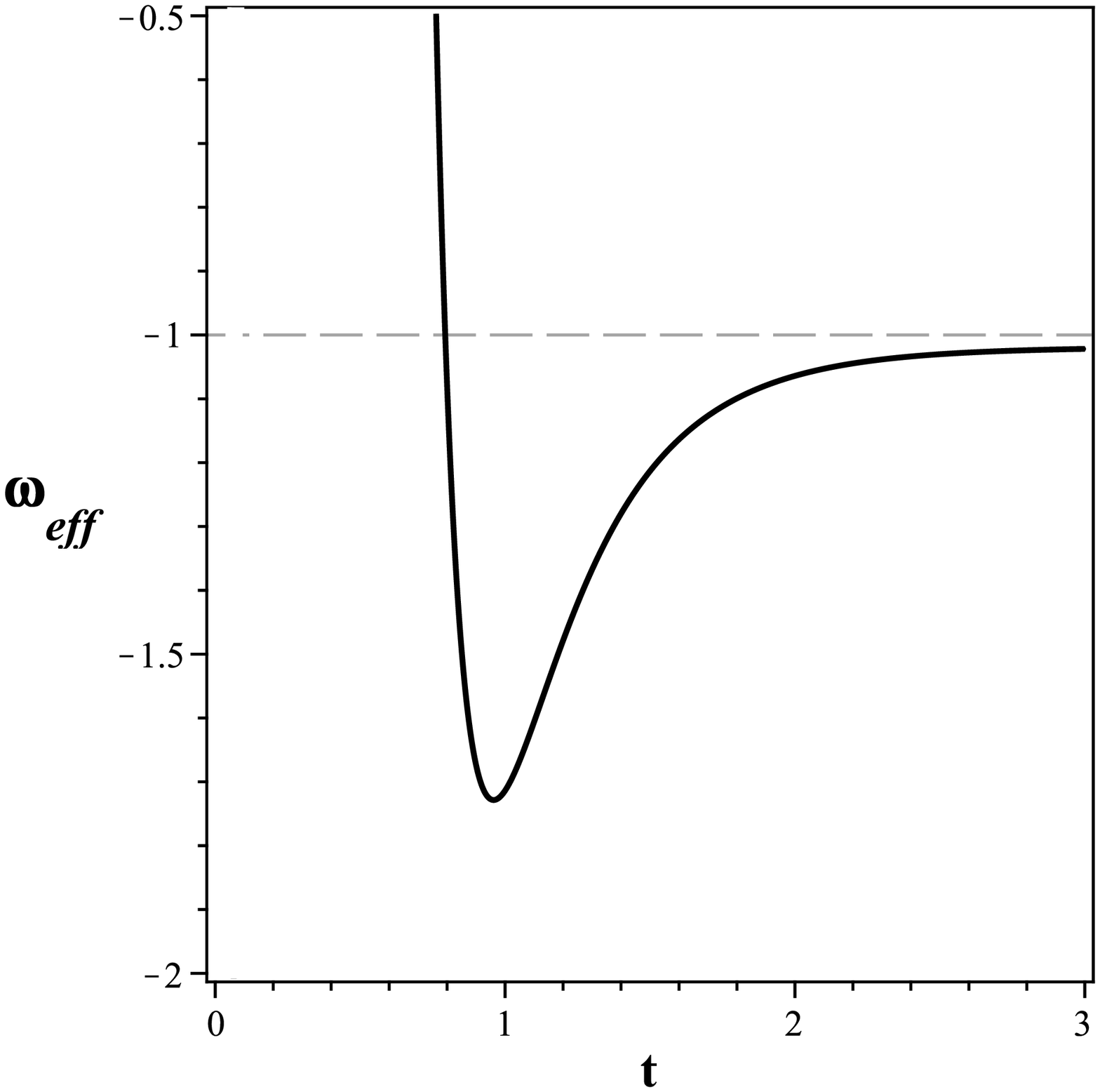}\\
Fig. 3: \, The graph of $\omega_{eff}$,  plotted as function of time for linear and quadratic forms:\\
\bf{Left panel:} $F(R,\phi)=P_1(\phi)R+P_0(\phi)$, \bf{Right panel:} $F(R,\phi)=P_2(\phi)R^2+P_1(\phi)R+P_0(\phi)$,\\
$P_2(\phi)=\ln(\phi)$, $P_1(\phi)=\phi^\alpha$ and $P_0(\phi)=e^{\beta \phi}$. In both, $\alpha = -0.125$, $ \beta= -0.5$ and $\phi(0)=0.5$.\\
\end{tabular*}\\

Similar to the difficulties arise in numerically solving the forth order nonlinear differential equations in $F(R)$ gravity for the
scale factor $a(t)$ \cite{Capozziello1}, here in $F(R,\phi)$, even in relatively simple cases of linear, quadratic and cubic forms of the model, a large uncertainties on the boundary conditions have to be set to find out the scale factor. In a different approach in $F(R)$ gravity, the authors in \cite{Capozziello1}, instead of choosing
a parameterized expression for $F(R)$ and numerically solving the equations with the given boundary
conditions, relate the present day values of $F(R)$
derivatives to the cosmographic parameters (i.e. the deceleration $q_0$, the jerk $j_0$, the snap $s_0$ and the lerk $l_0$ parameters)
in order to constrain them in a model independent way
and suggest the kind of $F(R)$ theory could be able to fit the observed Hubble diagram. A similar approach which is beyond the scope of this work can be applied in our $F(R,\phi)$ gravity model, in a separate study.

\section{Cosmological Tests}

\subsection{Cosmological Redshift-Drift Test(CRD)}

In this section, to detect the cosmic acceleration more directly, following \cite{Lis}, \cite{Salehi}--\cite{Liske}, we employ the time derivative of the redshift in terms of  the Hubble parameter as
\begin{equation}\label{zdot2}
 \dot{z}=H_{0}[1+z-\frac{H(z)}{H_{0}}],
\end{equation}
where $\dot{z}$ measures the rate of
expansion of the universe: $\dot{z} > 0 $ ($< 0 $) indicates the
accelerated (decelerated) expansion of the universe, respectively. The redshift variation is
related to the apparent velocity drift of the source:
\begin{equation}\label{dv}
 \dot{v}=c\frac{\dot{z}}{1+z},
\end{equation}
where $\dot{v}=\frac{\Delta v}{\Delta t_{obs}}$ and $H_{0} =100 h Km/sec/Mpc $ \cite{McVittie}. Since the detection of the signals correspond to the redshift change and consequently velocity drift is very weak, to measure
the signals, observation of
the $LY\alpha$ forest in the QSO spectrum \cite{Jain}--\cite{Loeb} for a decade might be needed . In near future this requirement can be provided by a new generation of Extremely Large Telescope (ELT) which are equipped with aspectrograph to measure such
a small cosmic signal \cite{Cristiani}\cite{Pasquin,Pasquini}.

In the following, we use three sets of data (8 points) for redshift drift generated by performed Monte Carlo. We first revise the CPL parametrization model and  compare the $F(R,\phi)$ model with CPL model and observational data \cite{Lis}, \cite{Pasquin}--\cite{Liske}.

{\bf CPL model}

One popular parametrization, which explains evolution
of dark energy is the CPL model in which in a flat universe the time varying EoS parameter in
term of redshift $z$ is parameterized by,
\begin{equation}\label{CPL1}
\omega(z)=\omega_{0}+\omega_{1}(\frac{z}{1+z}).
\end{equation}
In the model, the Hubble parameter also is given by
\begin{equation}\label{CPL2}
\frac{H(z)^{2}}{H^{2}_{0}}=\Omega_{m}(1+z)^{3}+(1-\Omega_{m})e^{3\int^{z}_{0}\frac{1+\omega(\tilde{z})}{1+\tilde{z}}d\tilde{z}},
\end{equation}
where using Eq. (\ref{CPL1}) we can obtain the following equation for Hubble parameter
\begin{equation}
\frac{H(z)^{2}}{H^{2}_{0}} =\Omega_m(1+z)^3+(1-\Omega_m)(1+z)^{3(1+\omega_0+\omega_1)}\times  \exp{\left[-3\omega_1(\frac{z}{1+z})\right]}\cdot \label{Hr}
 \end{equation}
In CPL model the parametrization is fitted for different values of $\omega_0$ and $\omega_1$. The velocity drift with respect to the source
redshift is shown in Fig. (5) \cite{Lis}. A comparison of the model with the observational data shows that the best fit values are for $\omega_0=-2.2$ and $\omega_1=3.5$.

{\bf $F(R,\phi)$ Model}

From numerical calculation, in Fig. (4), it can be seen that for three linear, quadratic and cubic form of $F(R,\phi)$, the EoS parameter crosses $-1$ at different $z$. In linear case it crosses twice in the past while in quadratic and cubic cases it crosses only once. It shows that for cubic form, the phantom crossing is more appropriate with the observational data \cite{Sur} as sometimes in the past the EoS parameter becomes zero and positive while this is not the case in linear and quadratic forms. In Cubic case, the EoS parameter crosses the boundary at about $z=1.4$ and its present value is about $-1.3$\\

\begin{tabular*}{2.5 cm}{cc}
\includegraphics[scale=.25]{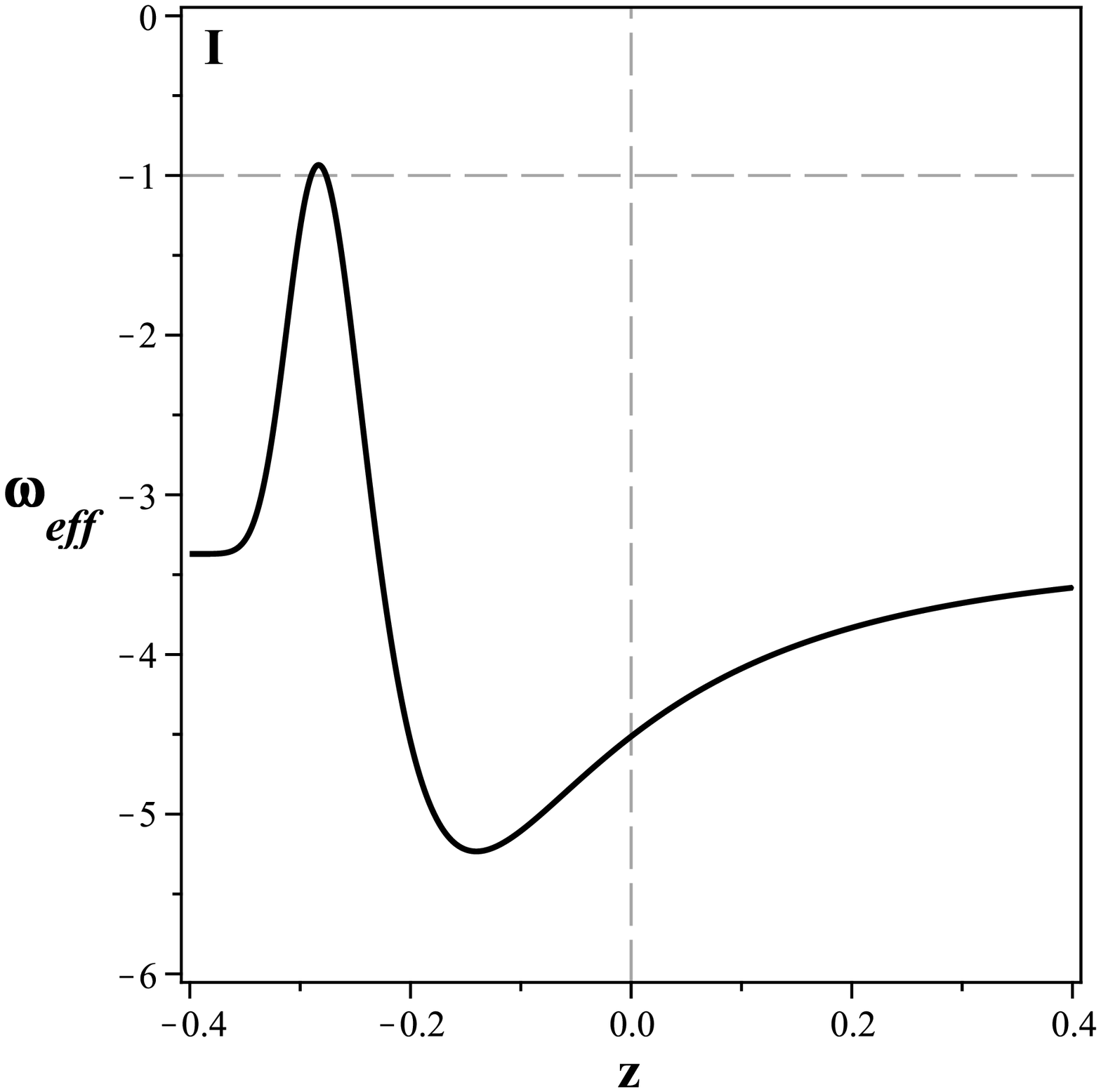}\hspace{0.2 cm}\includegraphics[scale=.25]{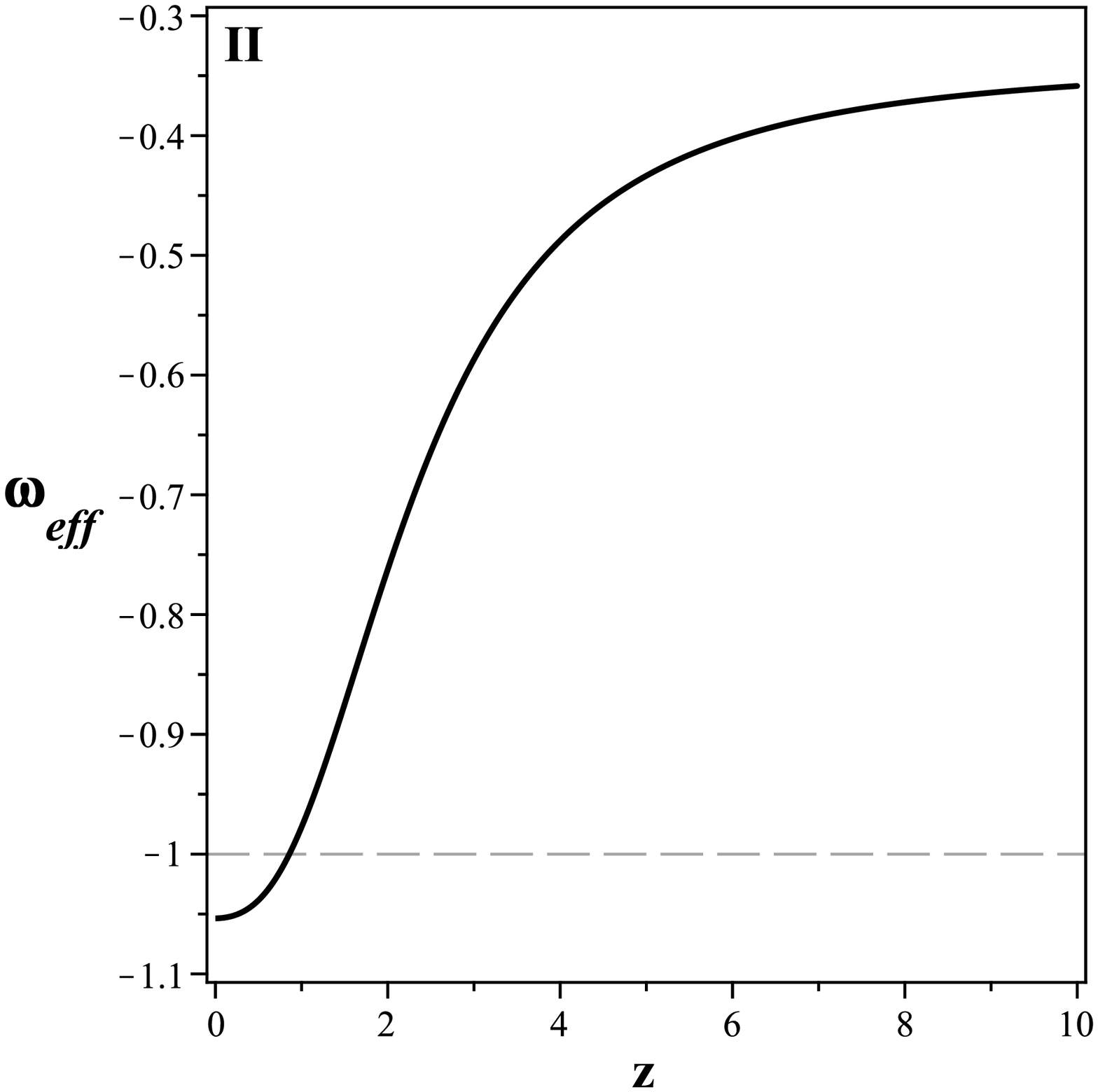}\hspace{0.2 cm}\includegraphics[scale=.25]{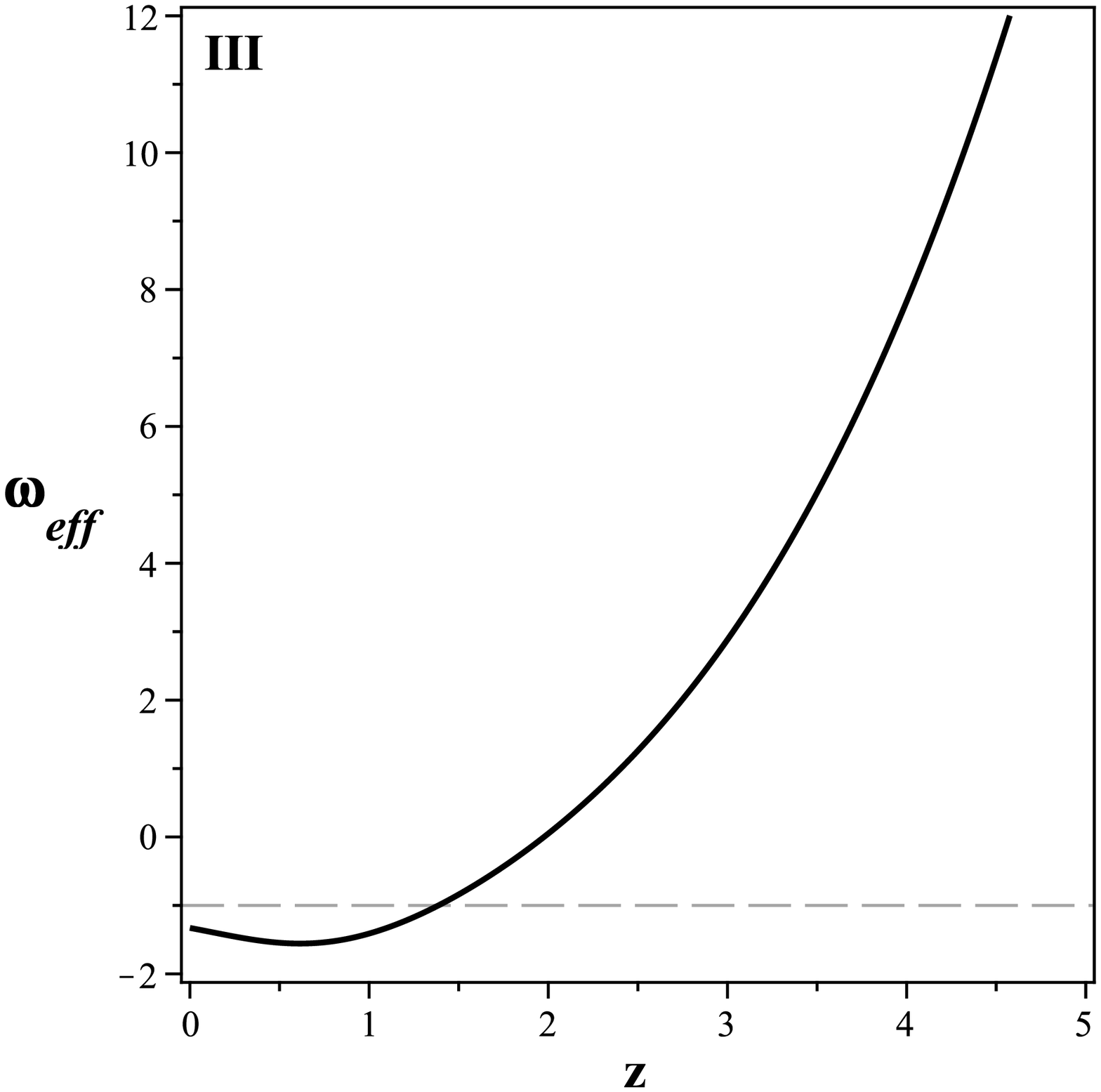}\\
Fig. 4: \, The graphs of $\omega$,  plotted as function of redshift $z$. \\
\textbf{I:} $F(R,\phi)=P_1(\phi)R+P_0(\phi)$, $\phi(0)=0.05$, \textbf{II:} $F(R,\phi)=P_2(\phi)R^2+P_1(\phi)R+P_0(\phi)$, $\phi(0)=0.5$\\
\textbf{III:} $F(R,\phi)=P_3(\phi)R^3+P_2(\phi)R^2+P_1(\phi)R+P_0(\phi)$, $\phi(0)=0.5$. $P_3(\phi)=\sin(\phi)$, $P_2(\phi)=\ln(\phi)$,\\
$P_1(\phi)=\phi^\alpha$ and $P_0(\phi)=e^{\beta \phi}$. In all cases we assume $\alpha= -0.09$ and $\beta= -0.6$. \\
\end{tabular*}\\

From equations (\ref{f1}), (\ref{f2}) and using $\frac{dH(t)}{dt}=-(1+z)H(z)\frac{dH(z)}{dz}$, we obtain the following expression
for EoS parameter,
\begin{eqnarray}\label{omegaeff}
\omega_{eff}=-1+\frac{(1+z)\frac{dr}{dz}}{3r},
\end{eqnarray}
where $r= \frac{H^{2}}{H^{2}_{0}}$. Solving equation (\ref{omegaeff}) for $H(z)$ we obtain,
\begin{eqnarray}\label{hz1}
\Big[\frac{H(z)}{H_{0}}\Big]^{2}=\exp\Big[{3\int^{z}_{0}\frac{1+\omega_{eff}(\tilde{z})}{1+\tilde{z}}d\tilde{z}}\Big]\cdot
\end{eqnarray}
 where from numerical computation in our model we find $\omega_{eff}$, and then by substitution into equation (\ref{hz1}) one obtains $H(z)$ for our model. Then, using equation (\ref{dv}), one calculates the velocity drift. Figure (5) shows the velocity drift against $z$ for linear, quadratic and cubic $F(R,\phi)$. The figure also display the parameterized CPL model result \cite{Lis1}.
In comparison with the CPL model and experimental data, our model is in a better agreement for cubic form of $F(R,\phi)$.\\

\begin{tabular*}{2.5 cm}{cc}
\includegraphics[scale=.4]{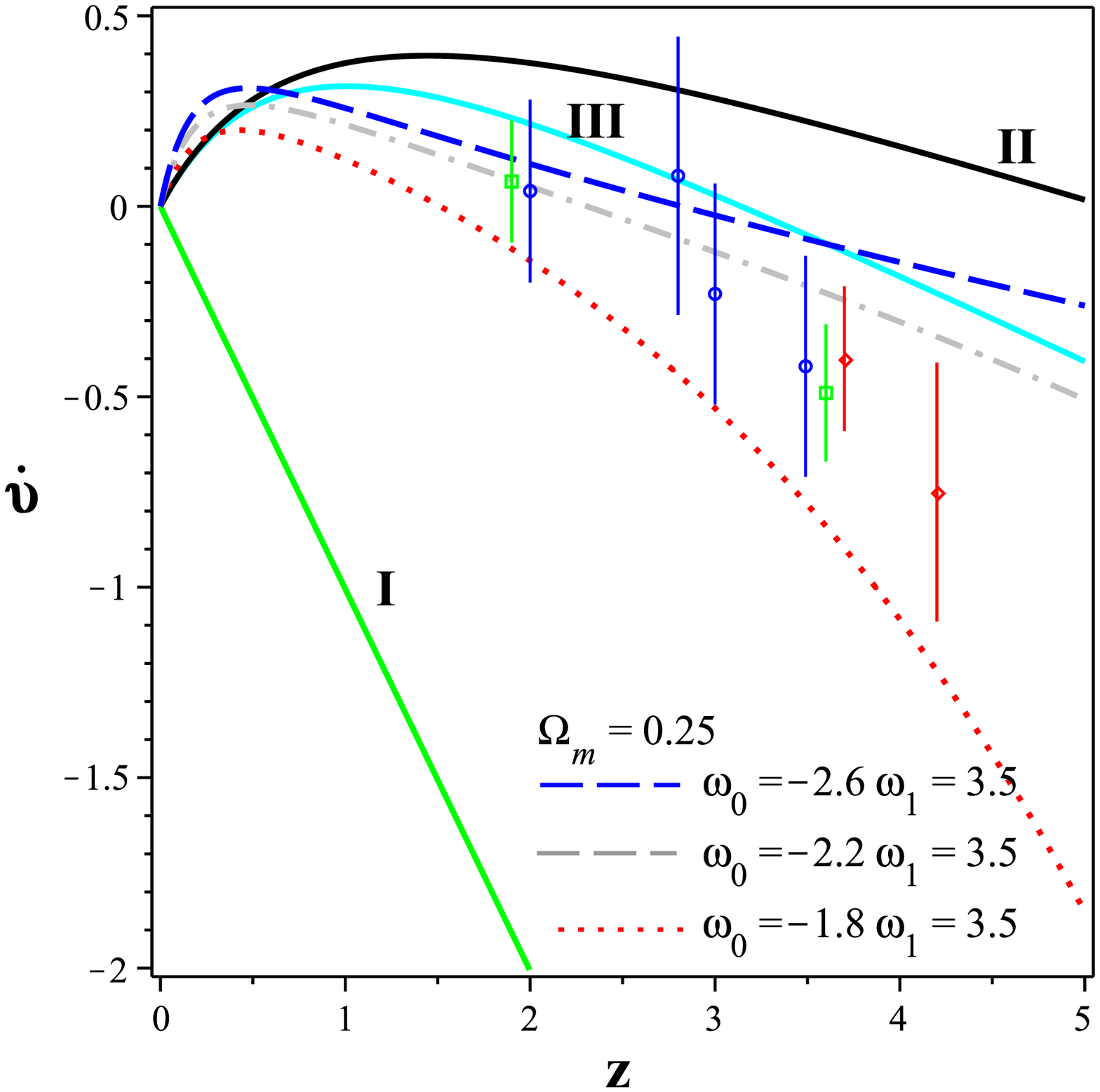}\hspace{0.2 cm}\\
Fig. 5: \, The graph of $\dot{v}$,  plotted as function of redshift $z$.\\
\textbf{I:} $F(R,\phi)=P_1(\phi)R+P_0(\phi)$, $\phi(0)=0.05$, \textbf{II:} $F(R,\phi)=P_2(\phi)R^2+P_1(\phi)R+P_0(\phi)$, $\phi(0)=0.5$\\
\textbf{III:} $F(R,\phi)=P_3(\phi)R^3+P_2(\phi)R^2+P_1(\phi)R+P_0(\phi)$, $\phi(0)=0.5$. $P_3(\phi)=\sin(\phi)$, $P_2(\phi)=\ln(\phi)$,\\
$P_1(\phi)=\phi^\alpha$ and $P_0(\phi)=e^{\beta \phi}$. In all cases we assume $\alpha= -0.09$ and $\beta= -0.6$. \\
\end{tabular*}\\
\\

\subsection{Luminosity $d_L(z)$ and the difference in the distance modulus $\mu(z)$}

Luminosity distance quantity, $d_L(z)$, determines dark energy density from observations. In CPL model with $H(z)$ given by (\ref{Hr}), it is in the form of,
\begin{equation}\label{dl}
d_{L}(z)=\frac{1+z}{H_0}\int_0^z{\frac{dz'}{\sqrt{\Omega_{m}(1+z')^{3}+(1-\Omega_{m})(1+z')^{3(1+\omega_{0}
 +\omega_{1})}\times e^{\left[-3\omega_{1}(\frac{z'}{1+z'})\right]}}}},
 \end{equation}
where $\Omega_m$ is the matter density. In addition, the difference $\mu(z)$ between the absolute and apparent luminosity of a distance object is given by, $\mu(z)=25+5log_{10}d_L(z)$.
In Fig. 6, we compare our model with CPL parametrization model for luminosity distance and with the observational data for  $\mu(z)$. It shows that the cubic form of our model is in a better agreement with the CPL model for luminosity. It also shows that again cubic form of the model fits the observational data.\\

\begin{tabular*}{2.5 cm}{cc}
\includegraphics[scale=.35]{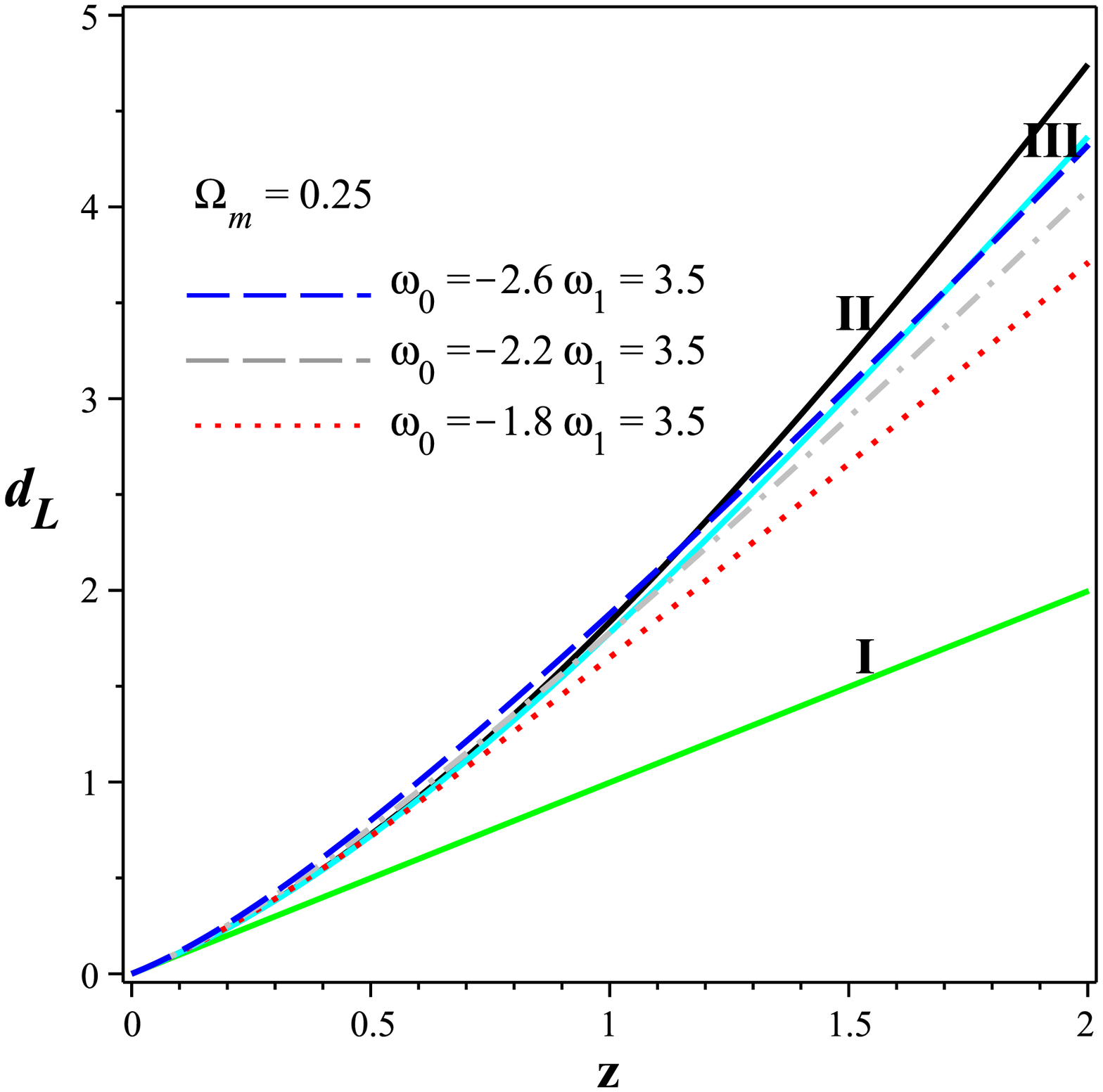}\hspace{0.5 cm}\includegraphics[scale=.35]{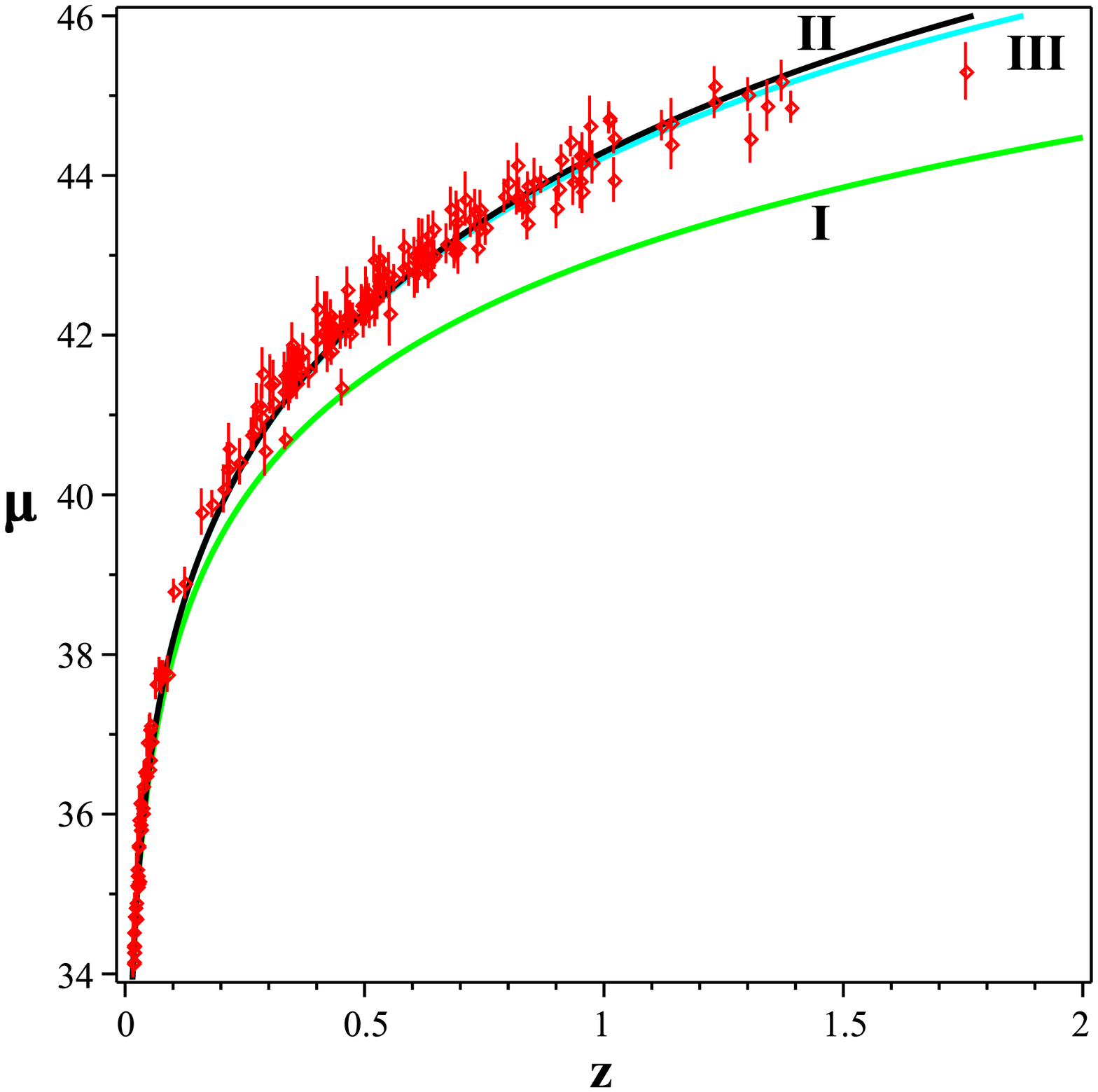}\hspace{0.2 cm}\\
Fig. 6: \, The graph of luminosity $d_L$ and the difference in the distance modulus $\mu$,\\  plotted as function of $z$
 for CPL parametrization and observational data for our model. \\
\textbf{I:} $F(R,\phi)=P_1(\phi)R+P_0(\phi)$, $\phi(0)=0.05$, \textbf{II:} $F(R,\phi)=P_2(\phi)R^2+P_1(\phi)R+P_0(\phi)$, $\phi(0)=0.5$\\
\textbf{III:} $F(R,\phi)=P_3(\phi)R^3+P_2(\phi)R^2+P_1(\phi)R+P_0(\phi)$, $\phi(0)=0.5$. $P_3(\phi)=\sin(\phi)$, $P_2(\phi)=\ln(\phi)$,\\
$P_1(\phi)=\phi^\alpha$ and $P_0(\phi)=e^{\beta \phi}$. In all cases we assume $\alpha= -0.09$ and $\beta= -0.6$. \\
\end{tabular*}\\

\section{Conclusion}

In this paper, we study the evolution of the gravitational and scalar fields in modified gravity in which a reconstructed $F(R,\phi)$ as a polynomial function of $R$ of degree $n$ is introduced. We find that the behavior of the scale factor of the universe depends on the degree of the polynomial.
For linear $F(R,\phi)$ a bouncing behavior and for a quadratic one an inflection behavior for the universe is shown. The evolution of the cosmological EoS parameter against time in both linear and quadratic cases, shows a crossing of the phantom divide line at $t>0$ in the past with a transition start from $\omega_{eff} <-1$ to the $\omega_{eff}>-1$ in linear case and from $\omega_{eff} >-1$ to the $\omega_{eff}<-1$ in quadratic case. A comparison of the phantom crossing in terms of redshift $z$ for linear, quadratic and cubic forms are given.

With respect to the observational data it shows that the quadratic and cubic forms are in better agreement with the data. It shows that in both cases the current value of the EoS parameter is between $-1$ and $-1.4$, the crossing occurs at about $z=0.5$ for quadratic and about $z=1.4$ for cubic case, and transition from matter to dark energy dominated era occurs only in cubic case

We then analyze the model with respect to the CPL model and observational data. The first cosmological test is the CRD test. The variation of
velocity drift for linear, quadratic and cubic forms of the polynomial with redshift
$z$ is shown in Fig. (5). Checking between our model with CPL parametrization model and observational data shows that cubic form is in better agreement with the experimental data. The second cosmological test is luminosity distance $d_L$ and the difference in the distance modulus $\mu$ that the former is calculated in comparison with the CPL parametrization model and the later with respect to the observational data. Again, for luminosity distance it shows that the the cubic from of our model is in better agreement with the CPL model. For $\mu(z)$, it show that again the cubic form of the model very well fit with the observational data. Although, due to numerically difficult computation we have not considered the higher order polynomial forms of $F(R,\phi)$, but from these three linear, quadratic and cubic forms we may conclude that the presence of the higher order terms in the model better fits the model with the observational data.

\end{document}